\documentclass[aps,prl,superscriptaddress,showpacs,floatfix,twocolumn]{revtex4}

\usepackage{graphicx}
\usepackage{bm}
\begin{document}

\title{``Jet-Ridge'' effect in heavy ion collisions as a back splash from stopped parton}
\author{V.~S.~Pantuev\\
University at Stony Brook, Stony Brook, NY 11794-3800
}
\altaffiliation[on leave from ]{Institute for Nuclear Research, Russian Academy of Sciences, Moscow, Russia}
\begin{abstract}
I propose a simple explanation of the ``ridge'' seen in the near-side jet two-particle correlation function 
in heavy ion collisions at RHIC. This could be a cumulative shock wave produced in 
liquid-like matter by an energetic stopped parton. Splash of liquid in the direction opposite 
to the fast penetrating object is a known effect. In nucleus-nucleus collision the trigger 
is an escaped non-interacting parton. Partner parton could be stopped in the 
medium forming a conical-like shock wave with density depletion behind the stopped 
parton. 
In the proposed scenario shock waves will move in three directions: two waves will form  
a correlation structure at angles differ 
from $\phi = \pi$ and are usually called a Mach Cone. The other direction is exactly opposite 
to the original direction of stopping parton. 
The wide rapidity distribution of the ridge is caused by the rapidity swing of the 
away-side jet and longitudinal expansion of the system along the beam direction. 
The calculation of such a  
shock wave is very complicated, but some phenomenological observables could be 
explained and predicted.

\end{abstract}



\maketitle
\begin{figure}[thb]
\includegraphics[width=1.0\linewidth]{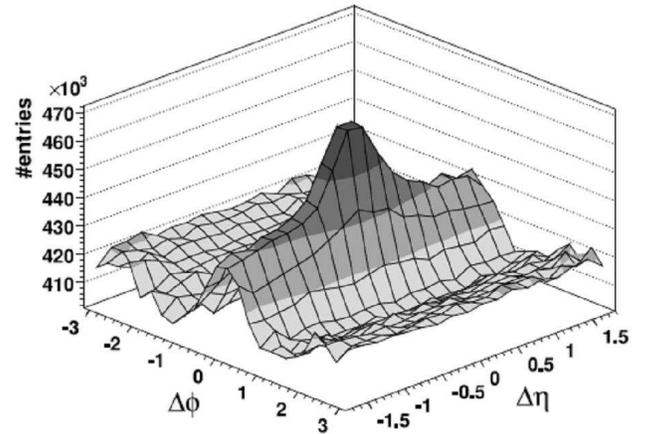}
\caption{\label{fig:ridge} The raw dihadron correlation function in azimuthal angle $\Delta \phi$ 
and pseudorapidity $\Delta \eta$ in central Au+Au collisions for trigger particle with 
transverse momentum between 3 and 4 GeV/c and associated particle with momentum 
above 2 GeV/c~\cite{star_ridge}.}
\end{figure}

Two-particle azimuthal correlation studies are a powerful tool to investigate jet production in 
relativistic nucleus-nucleus collisions. First, the STAR collaboration  observed  
jet-like structures~\cite{star_jet}. Then, the PHENIX 
collaboration  found 
modifications of di-jet hadron pair correlations in Au+Au collisions~\cite{mach_cone}, usually attributed to 
Mach Cone formation. Recently an enhanced near-side correlated yield within a large rapidity 
range to the trigger particle~\cite{star_ridge}, called as the {\it ridge}, Fig.~\ref{fig:ridge}, 
was observed by the STAR collaboration. 
In Fig.~\ref{fig:ridge} one can see a clear jet fragmentation peak at $\Delta \phi$=0 and $\Delta \eta$=0. 
In addition to the jet peak and some elliptic flow modulated background,  the {\it ridge}, 
a prominent enhancement in wide 
$\Delta \eta$ rapidity range is seen. Within the sensitivity of the STAR measurement, 
$\pm$1.5 units in pseudorapidity $\Delta \eta$, the ridge-like correlation is almost uniform. The ridge yield 
significantly increases with centrality, but is almost independent of trigger particle momentum. 
The momentum spectrum of secondary particles associated with the ridge is very close to the inclusive 
particle spectrum, but is slightly harder with a slope difference $\Delta T \approx$40-50 MeV.

\begin{figure*}[hbt]
\includegraphics[width=0.9\linewidth]{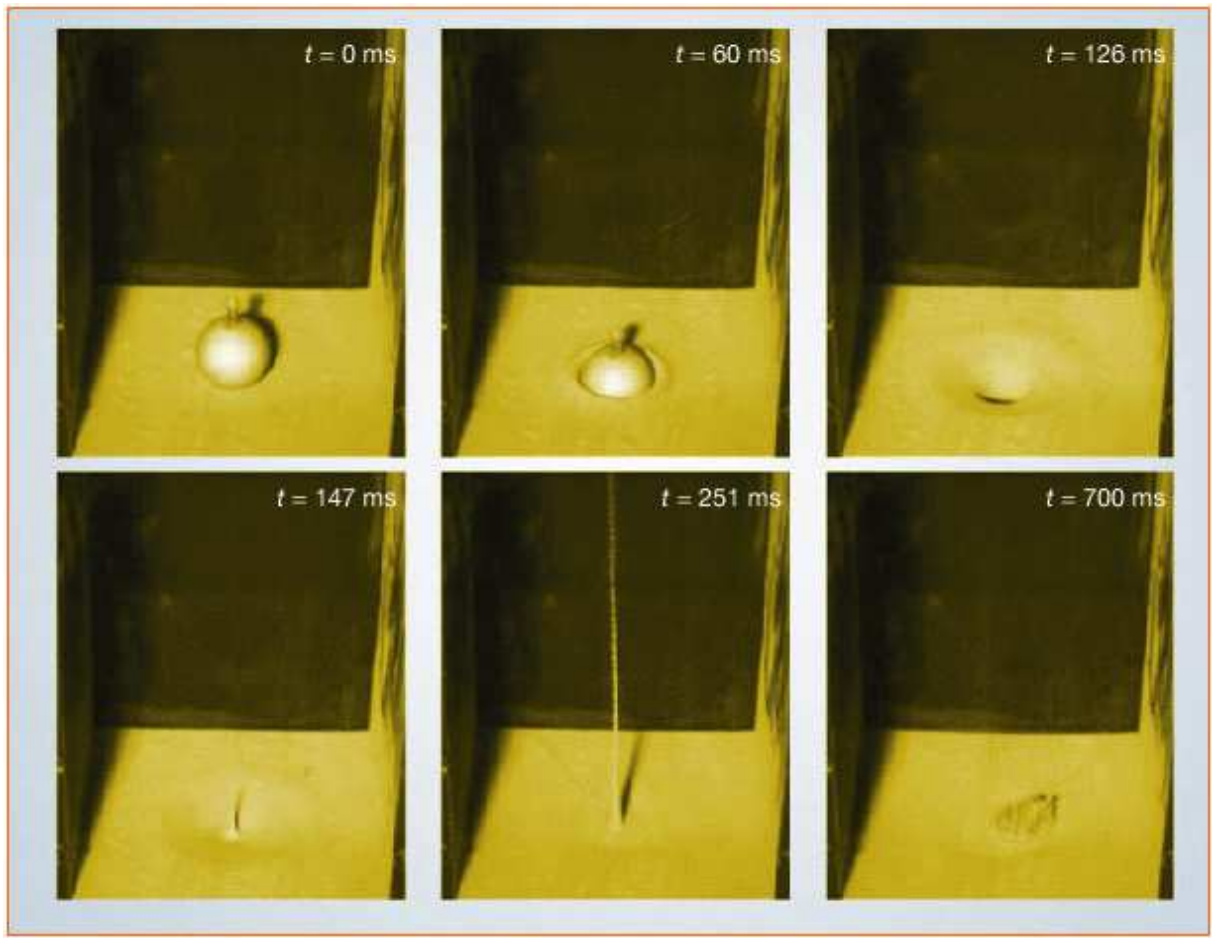}
\caption{\label{fig:sand_jet}Very fine sand is flowing by air through it. 
At time t=0 ms, the heavy ball is released and immediately starts to sink into the sand; at t=130 ms, 
a sand jet emerges, which reaches its final height at t=180 ms $\lbrack$4$\rbrack$. 
}
\end{figure*}

Without going into the details of various explanations of the ridge proposed so far, 
I'll  
start with a useful illustration. Fig.~\ref{fig:sand_jet} taken from an article in the 
journal {\bf Nature}~\cite{sand} shows a nice 
sand jet formed after the impact of a heavy object. Searching on the Web you can find many 
similar interesting photographs with liquids and drops.

This illustration shows how some portion of the bulk matter can be radiated opposite 
to the impact direction. In heavy ion collisions at RHIC, where the produced nuclear matter 
looks more like a liquid and strongly absorbs energetic partons~\cite{white_paper}, very similar 
effects should persist. If we trigger our apparatus on the particle escaped from the medium, 
and if the other parton, produced in the opposite direction, is stopped by the medium, there 
should be a splash of the bulk matter in response to the absorbed parton.

\begin{figure*}[hbt]
\includegraphics[width=0.9\linewidth]{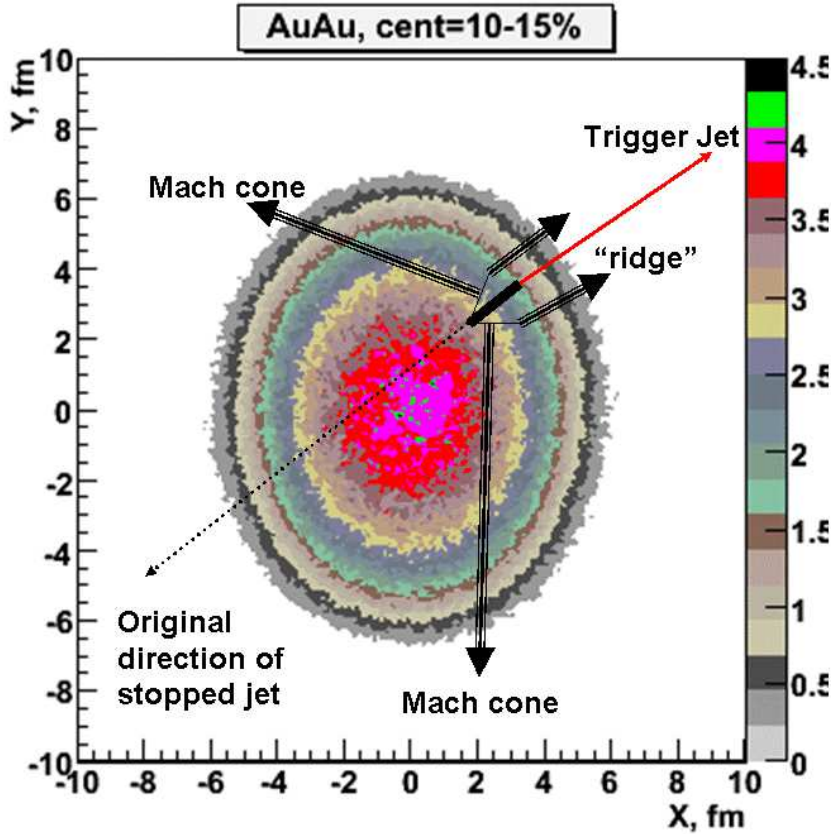}
\caption{\label{fig:ridge_details}Plotted here is the transverse plane profile of two interacting 
gold nuclei at centrality 10-15\%. Different colors correspond to different participant nucleon density. 
The impact parameter vector is oriented along the x-axis, and the beam direction is 
perpendicular to the figure plane. 
The thin arrow going to the up right illustrates  
the direction of the trigger jet, which escapes from the interaction zone. The dotted arrow going in 
the opposite direction shows the original direction of the absorbed parton. The thick black line between 
these two arrows demonstrate an effective path length of about 2 fm before the parton was 
absorbed. The directions of the Mach cone fronts are also shown. The matter between two short arrows, 
pointing in about 
the same direction as the trigger jet, is the portion of the nuclei interaction region, which will be
splashed out and will form the ridge.
}
\end{figure*}

Fig.~\ref{fig:ridge_details} schematically illustrates the geometry of the collision,   
Mach cone, and ridge formations. 
Two initial energetic partons interact and fly in opposite directions. One parton escapes
from the interaction region and fragments in vacuum producing the trigger particle. 
The other parton travels a short distance and then is stopped by the medium. Supersonic 
motion of this parton will produce internal volume tension, which will be released as 
Mach cone fronts at some angle and as a cumulative superposition of the waves in opposite direction. 
The latter waves will splash out some portion of the bulk matter forming the ridge. 
Because of a strong parton absorption in the core of the interacting zone, the trigger 
particle will be surface biased. The stopped parton also can not penetrate too deep 
into the medium. The penatration distance should be on the order of 2-3 fm~\cite{me} and no 
or very little depends on parton momentum. 
This surface bias will fix a portion of the matter splashed as 
the ridge per each absorbed parton, and  thus will not depend much on the centrality of the collision. 

I have to list some conclusions, explanations and predictions in case the back splash is the 
origin of the ridge:

1. The near side jet (trigger particle) is not the source of the ridge. It serves only as a trigger  
for this type of event.

2. The stopped parton/jet is the source of the ridge.

3. The Mach cone formation and the ridge are effects of the same cause.

4. The surface bias and momentum independence of the parton absorption makes the ridge independent 
of parton momentum. This is observed experimentally~\cite{star_ridge}. 

5. The rapidity distribution of the ridge should be wide, not smaller then the swing of away-side 
jet in rapidity, which is about $\pm$1. Plus longitudinal expansion will make it even wider. 
Experimentally indeed we see a wide rapidity distribution~\cite{star_ridge}.

6. The ridge should exhibit very similar properties as the bulk matter, but

7. Because of some velocity boost, the particle spectrum from the ridge should be harder than 
the inclusive spectrum. Also observed experimentally~\cite{star_ridge}. 

8. This velocity boost will change the heavy particle spectrum more than for 
light pions. This is a prediction.

9. Particle composition of the ridge is the same as in the bulk. 

\begin{figure}[thb]
\includegraphics[width=1.0\linewidth]{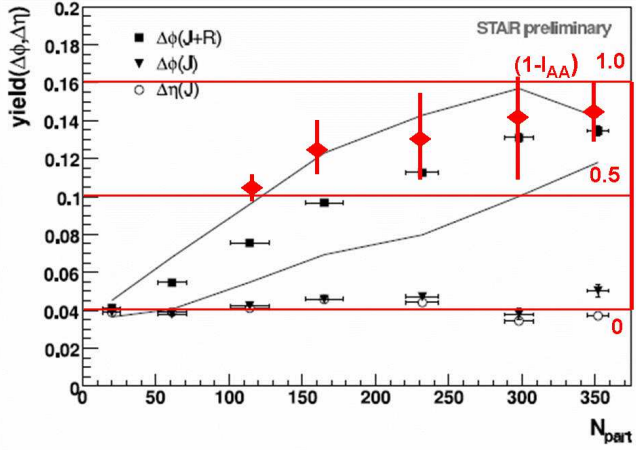}
\caption{\label{fig:yield} Black squares are data for near-side yield for associated particles 
for the sum ridge+jet, ;  triangles -- for jet defined in $\Delta \phi$ direction; open symbols --
from jet defined in $\Delta \eta$ direction~\cite{star_ridge}. On the right, the vertical axis 
represent  $1-I_{AA}$ values, which are plotted as red diamonds with statistical error
bars. Large systematic errors are not show. These points are plotted in 
rather schematic manner. Data for  $I_{AA}$ are taken from~\cite{star_jet}. }
\end{figure}

10. The ridge yield should depend on probability of absorbing the away-side jet. In other words, 
the ridge yield is proportional to $1-I_{AA}$, where $I_{AA}$ is 
the probability to observe away-side jet in A+A collisions compared to $p+p$ collisions. 
To check this statement, in Fig.~\ref{fig:yield} we present the ridge yield data alone with 
scaled arbitrary $1-I_{AA}$ values. Data for  $I_{AA}$ are taken 
from the first STAR jet measurements~\cite{star_jet}. Large error bars preclude a 
definite conclusion, but the tendencies between the ridge yield 
and $1-I_{AA}$ are similar.

11. Because the Mach cone and the ridge share their origin, these events should have 
an additional azimuthal distribution Fourier 
component, $v_3$, decoupled from the ``standard'' $v_2$ and $v_4$  defined by the 
reaction plane orientation.

12. For trigger hardon with relatively low momentum, less than 2-3 GeV/c, events could bemay have been 
triggered by the ridge (or Mach cone) particle, not to the survived jet. Indeed, two particle 
correlations at transverse momentum as low as 0.2-0.4 GeV/c exsibit some structure at 
$\Delta \phi \approx$2/3$\pi$~\cite{mitchell}. 

13. Hard direct photon-hadron correlations should also demonstrate the near side ridge and 
Mach cone.

14. Each absorbed parton produces a triple structure in the $\phi$ direction: two directed 
by Mach cone and one by the ridge with an angle of about 2/3$\pi$ between them. 
If two hard partons were produced and absorbed deep inside the bulk matter, 
searching in wide rapidity range it will be possible  
to find two triple structures, which form six relatively symmetric directions at 1/3$\pi$ 
angles. However, it is difficult to trigger for such event, and the reaction plane orientation with 
significantly distorts the six-directed texture.

It is very difficult to justify theoretically the proposed scenario, but experimental 
confirmation that the Mach cone structure and the ridge are from the same origin will 
be a definite proof of liquid-like structure of the produced matter.

\begin{acknowledgments}
This work was partially supported by the US-DOE grant DE-FG02-96ER40988. 

\end{acknowledgments}

\end{document}